\documentstyle[12pt]{article}
\title{
ON TEMPERATURE {\em versus} DOPING PHASE DIAGRAM OF HIGH CRITICAL
TEMPERATURE SUPERCONDUCTORS
}
\author{{\sl V.M.~Loktev}\\
{\sl  Bogolyubov Institute for Theoretical Physics,}\\
{\sl Metrologichna str. 14-b, Kyiv, 252143 Ukraine}\\
{\sl V.M.~Turkowski}\\
{\sl Shevchenko Kyiv University}\\
{\sl Acad. Glushkova prosp. 6, Kyiv, 252127 Ukraine}}
\date{}
\begin{document} \maketitle


\begin{abstract}
The attempt to describe the bell-shape dependence of the critical
temperature of high-$T_{c}$ superconductors on charge carriers
density is made. Its linear increase in the region of small densities
(underdoped regime) is proposed to explain by the role of the order
parameter phase 2D fluctuations which become less at this density
growth. The critical temperature suppression in the region of large
carrier densities (overdoped regime) is connected with the appearance
(because of doping) of the essential damping of long-wave bosons
which in the frame of the model proposed define the mechanism of
indirect inter-fermion attraction.
\end{abstract}


\eject

\section{Introduction}

In spite of non-falling down the intensity of study of high-$T_{c}$
superconductors (HTSCs) the comprehension of many regularities which
define their physical behaviour is not still achieved.
First of all, it is put down to the HTSC normal state
properties, the description of which in the framework of standard
Fermi-liquid theory proves to be impossible if the carrier
concentration $n_{f}$ in corresponding samples is like that one that
the resulting critical temperature $T_{c}$ in them is less than some
characteristic for every compound optimal value $T_{c}^{max}$. The
latter in all (practically, without exclusion) HTSC copper oxides
appears because, as it is well-known (see, for example, the rewiev
\cite{Lok1}),
the conductivity (consequently - superconductivity) these, initially
quasi-2D antiferromagnetic insulators, results from their doping by donor (Nd, Pr)
or acceptor (Sr, Ba, O) ions.

Then the temperature $T_{c}$ being formed at increasing of carrier
density (or, what is the same, Fermi-energy) in a system also
grows up. This growth, however, is rather quickly stopped
and after some (not large upon $n_{f}$ values) part of "saturation"
the drop to zero of the function $T_{c}(n_{f})$ follows, which thus
acquires the bell-shape form. As a result, the HTSC compounds with
relatively large concentration of itinerant carriers
(so called, overdoped regime) become non-superconducting metals the
behaviour of
which to a certain degree can be carried out on the base of conventional
theory of Fermi-liquid \cite{Dess}. The corresponding carrier
concentration values for underdoped regime are such ones
that the $T_{c}(n_{f})$ increase is observed for $n_{f}$ from $\simeq
0.04-0.08$ carrier per $CuO_{2}$ layer cell (i.e. from the threshold
of the insulator-metal transition) to $T_{c}^{max}
\simeq T_{c}(0.15-0.18)$; the temperature $T_{c}(n_{f})$ becomes
zero when $n_{f}\simeq 0.25-0.28$ \cite{Keim,Tak,Nag}. If the initial
part of the function $T_{c}(n_{f})$, where
$dT_{c}(n_{f})/dn_{f}\simeq const>0$ \cite{Uem,Yam}, can be (at least
qualitatively) interpreted (see Refs.\cite{Emer,Gus,Abr} and the
review \cite{Lok4}) proceeding from the theory of the crossover from
the Bose-Einstein condensation of separate composite bosons (local
pairs) and their superfluidity to the superconductivity of the BCS
type, then the reason of $T_{c}(n_{f})$ suppression in the BCS
carrier concentration region remains, in fact, unknown yet.

From the very begining the above-mentioned behaviour of $T_{c}(n_{f})$
was ascribed to the strong electron-electron correlations
(see, for example, the book \cite{Plak}) and to the filling of
usually narrow Hubbard
subbands by itinerant carriers.  This question was intensively
investigated by Prof. I.V.~Stasyuk and his collaborators
\cite{St1,St2,St3,St4}.

However a little bit later
it becomes evident that carrier concentrations corresponding to
a disappearance of superconductivity are so small that such a filling
of the whole valent (in fact, conduction) band in
HTSCs is impossible. The alternative version of the HTSC effect
disappearance most consistently considered in
Refs.\cite{Fried,Mar} (see also \cite{Lok1,Dess}) consists in assumption
that bare Fermi-level in HTSCs proves to be in the vicinity of
the extended saddle point of the electronic spectrum.
According with this poin of view at $n_{f}$ growth the Fermi-energy
$\epsilon_{F}\simeq {\bf k}_{F}^{2}/2m$ (${\bf k}_{F}$ - Fermi
momentum, $m$ - effective mass) passes the van Hove singularity in
the conduction band density of states and, hence, $T_{c}(n_{f})$
reveals its maximum. If it would be so the function $T_{c}(n_{f})$,
being similar to BCS one, would have the exponential dependence on
$n_{f}$ far from
$T_{c}^{max}$ what, as it is seen from experiments and was already
said, is not confirmed by observations.  On the other hand, the very
appearance of the van Hove spectrum singularity requires fine-tuning
(for example, because of rather definite ratio between the hole
next-near and nearest hopping constants) what seems can be hardly
achieved even in one compound not speaking of many ones.

At the same time less attention was paid to the idea that the
weakness of superconductivity can be conditioned by "feedback"
effect, or the changes in the spectrum $\omega ({\bf k})$ of
intermediate bosons the exchange of which results in the fermion pairing.
These bosons in HTSCs can be attributed to phonons
(similarly to BCS-Eliashberg model) the role of which is actively
advocated by Ginzburg (see, for example, his rewiev \cite{Gin}),
magnons, or spin fluctuations, \cite{Kam,Pin}, quadrupole
$dd$-excitons (Gaididei-Loktev-Weber mechanism (see
Ref.\cite{Lok1})), plasmons \cite{Pash}, etc. It must be,
however, noted that if phonon spectrum does not (or very weakly)
depend on doping then long-wave excitations of magnetic subsystem
($dd$-excitons, as excitations over a magnetic background, can be
also considered here) are strongly suppressed because of the
long-range magnetic order destruction in metallic phase of HTSCs
\cite{Lok1,Plak}.

Namely spin waves are subjected the most appreciable doping effect;
their low-frequency region (for the wave vectors less than some
characteristic value $k_{min}$) acquires diffusion form,
or becomes overdamped.
In the insulating phase of HTSCs
as it is shown in Ref. \cite{Ivan}, $k_{min}\sim
n_{f}$; in their mettalic phase the magnon damping becomes even more
strong and $k_{min}\simeq 2k_{F}\sim n_{f}^{1/2}$ \cite{Lok2,Lok1},
what is completely in line with degradation of the correlation
magnetic length $\xi_{mag}\sim n_{f}^{-1/2}$ measured in HTSCs
\cite{Thur}.  In that way the long-wave damping $\gamma ({\bf k})$
increases up to so high values
($\gamma ({\bf k})>>\omega ({\bf k})$)
that corresponding intermediate bosons
(most probably, spin fluctuations) can not participate in interaction
transfer, "being out".

The solution of the self-consistent magneto-electronic problem
as a whole (i.e. the explicit solution of the equation of
superconductivity similar to the Eliashberg equation) is scarcely
possible now. Therefore below the attempt is made to consider
the simplest model with indirect iner-fermion attraction what
is provided by intermediate massive bosons with long-wave cut
which is proportional to $k_{F}$. The carriers to be supposed
appear in a system due to doping what corresponds to generally
accepted scenario of metallization of copper oxides. For the sake
of simplicity we shall omit any other dampings (in particular,
the carrier damping because of disorder the effect of which
was analysed, for example, in Ref.\cite{Lok3}).

Following to Refs.\cite{Gus,Lok4} we shall also suppose that in a
model 2D system (in fact, all HTSCs can be with a good accuracy
put down to this kind) the
superconducting condensate is formed in a way principally different
from the ordinary one. In such a case one has to differ the order
parameter formation temperature from the real (observable) critical
temperature in a sense that the absolute value of the former does
not become zero at and above $T_{c}$.

\section{Model and main equations}

The model Hamiltonian density of the electron-phonon
system can be written in the well-known form:
\begin{eqnarray} H(x) =-\psi_{\sigma}^{\dagger}(x)
      \left(\frac{\nabla^2}{2m}+\mu\right)\psi_{\sigma}(x) +
      H_{ph}(\varphi (x))\nonumber \\
      +g_{ph}\psi_{\sigma}^{\dagger}(x)\psi_{\sigma}(x)\varphi (x),
      \ \ \  x=\mbox{\bf
      r},t, \label{1}
 \end{eqnarray}
where $H_{ph}$ is the Hamiltonian of free phonons which
will be described more precisely below;
$\psi_{\sigma}(x)$
and $\varphi (x)$ are the fermion and boson field
operators, respectively;
$m$ is the effective mass of the fermi-particles,
$\sigma =\uparrow
,\downarrow $ - their spin variable, and
$g_{ph}$ is the electron-phonon
coupling constant; $\mu$ in (\ref{1}) is the chemical
potential of fermions, which fixes their average density
in a system; we put $\hbar=k_B=1$.

As it was mentioned in Introduction, we shall model
boson-exchange interaction taking into account the dependent on
doping saturation of long-wave bosons.
Under above mentioned assumptions about gradual non-participation of
the part of the bosons in an attracting interaction formation
the simplest way to describe that can be achieved by making use
of the free-phonon propagator in the form:
\begin{equation} D(\omega , {\bf k}) =-\frac{\omega^2
({\bf k})}{\omega^2-\omega^2 ({\bf k})+i\delta}\theta (k-k_{min}) ,
\  \  \  \delta\rightarrow 0,
\label{2}
\end{equation}
where
$\omega ({\bf k})$ is the boson dispersion law, and
$\theta (k)$ is the step function.
As it was pointed out above ${\bf k}_{min}$ in
(\ref{2}) is some characteristic wave vector which
devides the region of overdamped ($k<k_{min}$) and long-lived
($k>k_{min}$) intermediate bosons (here: phonons by definition).
Although magnetic correlation length measurements result in
$k_{min}=2k_{F}$ we shall adopt a more general (or soft) relation supposing
that $ k_{min}=\alpha k_{F}\equiv\alpha\sqrt{2m\epsilon_{F}}$
where $\alpha$ is some free-parameter.

It is very important that the Hamiltonian (\ref{1}) is
invariant with respect to the global symmetry transformations
\begin{equation}
\psi_{\sigma}(x)\rightarrow\psi_{\sigma}(x)e^{i\theta},\ \
   \psi_{\sigma}^{\dagger}(x)\rightarrow\psi_{\sigma}^{\dagger}(x)
   e^{-i\theta},\label{s1}
\end{equation}
which in 2D case (unlike to 3D one) remain unbroken and phase
transition is here accompanied by change in the correlation function
behaviour only.

The $T-n_{f}$ phase diagram of a system can
be calculated by using
the Hubbard-Stratonovich method, generalized on
the case of non-local (indirect) interaction
(so called auxiliary bilocal
field method).
For the finding of the grand partition function $Z$
it is useful to pass to Nambu spinors:
$
\Psi^{\dagger}=(\psi_{\uparrow}^{\dagger},
\psi_{\downarrow})$ and its conjugated one.
After that performing the integration
over bosonic fields it is easy to obtain the Lagrangian
\begin{eqnarray}
L=\Psi^{\dagger}(x)[-\partial_{\tau}+
(\frac{\nabla^2}{2m}+\mu )\tau_{z}]\Psi (x)\nonumber\\
-\frac{1}{2}\Psi (x_{1})\Psi^{\dagger}(y_{1})\tau_{z}
K(x_1,y_1;x_2,y_2)\Psi
(x_2)\Psi^{\dagger}(y_2)\tau_{z}\label{5}
\end{eqnarray}
of a system, where the integration over repeated indices
is supposed.
The kernel $K$ in (\ref{5}) will be defined below.

Let introduce the pairing order parameter
$$
\phi (x_1,y_1)=K(x_1,y_1;x_2,y_2)\Psi
(x_2)\Psi^{\dagger}(y_2)\tau_{z}
$$
\begin{equation}
\equiv\tau_{+}\phi (x_1,y_1)+
\tau_{-}\phi^{*}(x_1,y_1),
\label{6}
\end{equation}
where $\tau_{+}=\frac{1}{2}(\tau_{x}+i\tau_{y})$,
$\tau_{-}=\frac{1}{2}(\tau_{x}-i\tau_{y})$ (and $\tau_{z}$
in (\ref{5})) are the Pauli matrices.

Then adding to the Lagrangian $L$ a zero term
$$
\frac{1}{2}[\phi (x_1,y_1)-K(x_1,y_1;x'_1,y'_1)\Psi
(x'_1)\Psi^{\dagger}(y'_1)\tau_{z}]K^{-1}(x_1,y_1;x_2,y_2)
[\phi (x_2,y_2)
$$
$$
-K(x_2,y_2;x'_2,y'_2)\Psi (x'_2)\Psi^{\dagger} (y'_2)\tau_{z}]
$$
with the purpose to cancel the four-fermion interaction one comes
to the expression:
\begin{eqnarray}
L(x_1,y_1;x_2,y_2)=\Psi^{\dagger}(x_1)
[-\partial_{\tau}+(\frac{\nabla^2}{2m}+
\mu )\tau_{z}-\tau_{+}\phi (x_1,y_1)\nonumber \\
-\tau_{-}\phi^{*}(x_1,y_1)]\Psi (y_1)+
\frac{1}{2}\phi (x_1,y_1)K^{-1}(x_1,y_1;x_2,y_2)\phi
(x_2,y_2)\label{7} \end{eqnarray}
for the Lagrangian needed.
The Fourier
transformation of $K$ can be written as
$$
K(x_1,y_1;x_2,y_2)=
\int\frac{d^3Pd^3p_1d^3p_2}{(2\pi)^9}K_P(p_1;p_2)
exp[-iP(\frac{x_1+y_1}{2}-\frac{x_2+y_2}{2})
$$
$$
-ip_1(x_1-y_1)-ip_2(x_2-y_2)],
$$
($p_i=(\mbox{\bf p}_i,\omega_i)$ where $i=1,2$ and
$P=(\mbox{\bf P},\omega )$ are the relative and the centre
of mass momenta, respectively). Supposing now that
$K_P(p_1;p_2)$ is $P$-
independent we pass to the standard kernel form
\begin{equation} K(p_1;p_2)=g_{ph}^{2}D(p_1-p_2),\label{71}
\end{equation}
 which corresponds to the indirect inter-fermion interaction.

The partition function can be written then as
$$
Z=\int {\cal D}\Psi^{\dagger} {\cal D}\Psi {\cal D}\phi {\cal D}\phi
^{*} \exp\left[-\beta\int L(\Psi^{\dagger},\Psi ,\phi^{*},\phi
)dxdy\right]
$$
$$
\equiv \int{\cal D}\phi {\cal
D}\phi^{*}\exp(-\beta\Omega [{\cal G}]), \ \ \beta\equiv 1/T,
$$
In the last expression $\Omega [{\cal G}]$ is the thermodynamic
potential which in the "leading order" on $g_{ph}$ has the form
\begin{equation}
\beta\Omega [{\cal G}]=- \mbox{Tr} \left[\mbox{Ln}{\cal
G}^{-1}+\frac{1}{2}(\phi K^{-1}\phi )\right], \label{9}
\end{equation}
in which
\begin{equation}
{\cal G}^{-1}  =
-\left[\partial_{\tau}-\left(\frac{\nabla^2}{2m}+\mu\right)
\tau_{z}-\tau_{+}\phi-
\tau_{-}\phi^{*}\right]
\label{10}
\end{equation}
is the full fermion Green function.
After the direct minimization of
the potential (\ref{9})
it is easy to obtain the equation for auxiliary
$\phi$-field (or order parameter):
\begin{equation}
\delta\Omega /\delta\phi
=2\phi-tr\int\frac{d^2\mbox{\bf k}d\omega}{(2\pi
)^3}K(p;\mbox{\bf k},\omega ){\cal G}(\mbox{\bf k},\omega
)\tau_{x}=0.\label{11}
\end{equation}
Using (\ref{11}) one can readily arrive to
the well-known Cornwell-Jackiw-Tombou-\\
lis formula for the
effective action in the one-loop approximation \cite{Cor1}:  $$
\beta\Omega ({\cal G}) =-\mbox{Tr}\mbox{Ln}{\cal G}^{-1}
+\frac{1}{2}\mbox{Tr}{\cal G}K{\cal G},
$$
or, taking into account (\ref{10}) (or (\ref{11})),
\begin{equation}
\beta\Omega
({\cal G })=-\mbox{Tr}\left[\mbox{Ln}{\cal G}+ \frac{1}{2}({\cal
G}{\cal G}_0^{-1}-1)\right].\label{12} \end{equation}

To investigate the possibility of the
condensate formation in a 2D system it is convenient
in accordance with \cite{Gus} to pass to a modulus-phase
parametrization of the order parameter (cp. (\ref{s1}), where
$\theta =const$):
\begin{equation} \phi(x,y)=\rho (x,y)exp[-i(\theta (x)+\theta
(y))/2]\label{13} \end{equation} with simultaneous Nambu spinor
transformation \begin{equation}
\Psi^{\dagger}(x)=\chi^{\dagger}(x)exp[i\theta
(x)\tau_{z}/2],\label{141}
\end{equation}
corersponding to "separation" of the bare fermions on their
neutral $\chi (x)$ and charge $\theta (x)$ parts
(fermi- and bose-ones, respectively).

In the approximation that
$\rho (x,y)=\rho =const$ (see Ref.\cite{Lv}) and spatial
$\theta$-fluctuations are small one can (using
(\ref{13}) and (\ref{141})) obtain the next expressions
for ${\cal G}$ and $\Omega$ which are defined
in (\ref{10}) and (\ref{12}):
\begin{eqnarray}
{\cal G}^{-1} & = &
-\left[\partial_{\tau}-\tau_z\left(\frac{\nabla^2}{2m}+
\mu\right)+i\tau_{x}\rho \right.
\nonumber \\ &&
\left. -\tau_z\left(\partial_{\tau}\theta+
\frac{\nabla\theta^2}{2m}\right)-i
\left(\frac{\nabla^2\theta}{2m}+
\frac{\nabla\theta\nabla}{m}\right)\right]\nonumber \\ &&
\equiv G^{-1}(\rho )
-\Sigma (\partial\theta )  \label{15} \end{eqnarray}
and
$\Omega =\Omega_{kin}(\rho ,\nabla\theta )
+ \Omega_{pot}(\rho )$ with
$\Omega_{pot}(\rho )$ which is defined by (\ref{12}) at
$\nabla\theta =0$ and
$$
\beta\Omega_{kin}(\rho ,\nabla\theta )
=\mbox{Tr}[G\Sigma -G_0\Sigma+\frac{1}{2}G\Sigma
G\Sigma -
\frac{1}{2}G_0\Sigma G_0\Sigma
$$
\begin{equation}
+
\tau_{x}\frac{1}{2}i\rho G(G \Sigma + G \Sigma
G \Sigma)]=
\frac{T}{2}\int_{0}^{\beta}d\tau\int d^2\mbox{\bf r} J(\mu, T,
\rho (\mu ,T))(\nabla\theta)^2, \label{17}
\end{equation}
where the effective neutral fermion stiffness
$$ J(\mu ,T,\rho (\mu ,T))
$$
$$
=\frac{1}{8\pi}\left[\sqrt{\mu^2+\rho^2}+\mu +
2T\ln\left(1+\exp(-\frac{\sqrt{\mu^2+\rho^2
}}{T})\right)\right]
$$
\begin{equation}
-\frac{T}{4\pi}\left[1-
\frac{\rho^2}{4T^2} \frac{\partial}{\partial
(\rho^2/4T^2)}\right]\int_{-\mu
/2T}^{\infty}dx\frac{x+\mu /2T}
{\cosh^2\sqrt{x^2+\rho^2/4T^2}} \label{18}
\end{equation}
was introduced.

The evident analogy with XY-model
(two-component order parameter in 2D space)
gives
the equation for the temperature $T_{BKT}$ of the
Berezinskii-Kosterlitz-Thouless phase transition in a system,
namely (see Chapter 15 in the book \cite{Iz}):
\begin{equation} \frac{\pi}{2}J(\mu ,T_{BKT},\rho
(\mu ,T_{BKT})) = T_{BKT}.  \label{19}
\end{equation}
(Recall that the temperature $T_{BKT}$ plays the role of critical one
in 2D metals.)

The parameters $\mu$ and $\rho$ in (\ref{19}) being dependent of
$T$ are still unknown; therefore
it is necessary to obtain the equations which connect them
with carriers density $n_{f}$.
The first one follows from $(\ref{11})$ for $\rho\not= 0$:
\begin{equation}
1 =
T\sum_{m=-\infty}^{\infty}\int\frac{d^2\mbox{\bf
k}}{(2\pi)^2}
\frac{
K(\omega_m)}
{\omega_m^2+\xi^2
(\mbox{\bf k})+\rho^2},
\label{20}
\end{equation}
where $\omega_n=(2n+1)\pi T$ is the Matsubara fermion
frequencies, $\xi ({\bf k})={\bf k}^{2}/2m-\mu$ and
it was used the Einstein model for phonon dispersion:
$\omega ({\bf k})=\omega_{0}$. The dependence on
parameter $\alpha$ in Eq.(\ref{20}) is preserved through
the kernel $K$ (see (7)  and (2)).

The second one is defined by the condition
$V^{-1}\partial\Omega_{pot}(\rho )/\partial\mu =-n_{f}$ ($V$ is the
volume of a system) what results in the well-known number
equation:  \begin{equation} \sqrt{\mu^2+\rho^2}+\mu
+2T\ln\left[1+\exp
\left(-\frac{\sqrt{\mu^2+\rho^2}}{T}\right)\right]=
2\epsilon_F, \label{21}
\end{equation}
where the equality $\epsilon_F=\pi n_f/m$ was used;
it is correct for free 2D fermions with a quadratic dispersion law.

Thus, we have obtained a self-consistent set of the
Eqs.(17)-(19) needed to investigate the phase diagram of a
2D metal with arbitrary carrier density. The latter is the parameter
defined such a metal superconducting properties.

\section{Phase diagram of a system}

As it can be seen from the previous Section
(see also Refs.\cite{Gus,Lok4})
there exist
two characteristic temperatures in a system:
$T_{\rho}$, where formally the complete order parameter given by
Eq.(\ref{6}) arises but its phase is a random quantity, i.e.  $<\phi
(x,y)>=0$, and another one, $T_{BKT}<T_{\rho}$, where the phase of
the order parameter becomes ordered, so that $<\phi (x,y)>\not=0$.
It must be, however, stressed that the temperature $T_{\rho}$ is not a
real critical temperature; it only denotes the characteristic region
where the modulus of the order parameter achieves its maximal growth
at $T$ decreasing. Unlike $T_{\rho}$ the temperature $T_{BKT}$
does correspond to phase transition when correlators
$<exp[i\theta ({\bf r},\tau )]exp[i\theta ({\bf r}')]>$
as function of $|{\bf r}-{\bf r}'|$ change their behaviour \cite{Iz}.

Let us find the $n_{f}$-dependence of the temperatures
 $T_{\rho}$ and
$T_{BKT}$.
"Effective" temperature $T_{\rho}$ can be estimated
from the set (\ref{18})-(\ref{21}) in mean-field approximation
by putting
$\rho =0$.  Another temperature $T_{BKT}$ follows from the Eqs.
(17) and (19).

It is impossible to solve the equations obtained analytically,
so we shall do that by numerical calculation. Nevertheless,
some assymptotical expressions for these temperatures as
functions of $n_{f}$ can be found in
analytical form:

i) at $\epsilon_F/\omega_0\rightarrow 0$ one obtains
$T_{BKT}=\epsilon_F/8$,
and
$T_{\rho}$
satisfies the simple mean-field
equation $T_{\rho}\ln (T_{\rho}/\epsilon_F)= \omega_0\exp
(-2/\lambda)$, where $\lambda =g_{ph}^{2}m/2\pi$ is the dimensionless
coupling constant.

ii)
$\epsilon_{F}\rightarrow\epsilon_{F}^{cr}$;
the critical point $\epsilon_{F}=
\epsilon_{F}^{cr}$
(or $n_{f}=n_{f}^{cr}$, what is the same) at which
$T_{\rho}=T_{BKT}=0$
can be found from Eqs.(17)-(19). This unknown energy
is the solution of the equation
\begin{equation}
1=\frac{\lambda}{2}\ln\frac{(W-\epsilon_{F})[(\alpha
-1)\epsilon_{F}+\omega_{0}]}{
(\alpha
-1)\epsilon_{F}
(W-\epsilon_{F}+\omega_{0})}\rightarrow
\frac{\lambda}{2}\ln\frac{(\alpha
-1)\epsilon_{F}+\omega_{0}}{
(\alpha
-1)\epsilon_{F}
}|_{W\rightarrow\infty}.\label{20}
\end{equation}
($W$ is the conduction bandwidth determined by evident condition:
$W={\bf k}_{B}^{2}/2m$, where
${\bf k}_{B}$ is the Brillouin wave vector).
In other words, it follows from (\ref{20}) that
because of the long waves phonons begin to quit the interaction
transfer, there exist
 (at $\alpha >1$ only), the point where both
 $T_{\rho}$ and $T_{BKT}$ temperatures become zero what means that
superconductivity (but not conductivity) is
suppressed.  Near this point this temperatures have the
next behaviour ($W\rightarrow\infty$):
$$
T_{\rho}=\omega_0/\ln\frac{4\epsilon_{F}^{cr}
[(\alpha-1)\epsilon_{F}^{cr}+\omega_{0}]}
{\lambda\omega_{0}(\epsilon_{F}^{cr}-\epsilon_{F})};
$$
$$
T_{BKT}=
\epsilon_{F}^{cr}
\left[\frac{9}{144}(\alpha
-1)^{4}\left(
\frac{\lambda\omega_{0}
(\epsilon_{F}^{cr}-\epsilon_{F})
}{
\epsilon_{F}^{cr}[(\alpha-1)\epsilon_{F}^{cr}+\omega_{0}]}\right)^{2}
\right]^{1/5}.
$$

The results of numerical calculations of Eqs. (17)-(19)
are shown on Fig.1.

\begin{figure}[htb]
\begin{centering}
\setlength{\unitlength}{0.240900pt}
\ifx\plotpoint\undefined\newsavebox{\plotpoint}\fi
\sbox{\plotpoint}{\rule[-0.200pt]{0.400pt}{0.400pt}}%
\special{em:linewidth 0.4pt}%
\begin{picture}(1200,720)(0,0)
\font\gnuplot=cmr10 at 10pt
\gnuplot
\put(220,113){\special{em:moveto}}
\put(1136,113){\special{em:lineto}}
\put(220,113){\special{em:moveto}}
\put(220,697){\special{em:lineto}}
\put(220,113){\special{em:moveto}}
\put(240,113){\special{em:lineto}}
\put(1136,113){\special{em:moveto}}
\put(1116,113){\special{em:lineto}}
\put(198,113){\makebox(0,0)[r]{0}}
\put(220,530){\special{em:moveto}}
\put(240,530){\special{em:lineto}}
\put(1136,530){\special{em:moveto}}
\put(1116,530){\special{em:lineto}}
\put(198,530){\makebox(0,0)[r]{1}}
\put(220,113){\special{em:moveto}}
\put(220,133){\special{em:lineto}}
\put(220,697){\special{em:moveto}}
\put(220,677){\special{em:lineto}}
\put(220,68){\makebox(0,0){0}}
\put(351,113){\special{em:moveto}}
\put(351,133){\special{em:lineto}}
\put(351,697){\special{em:moveto}}
\put(351,677){\special{em:lineto}}
\put(351,68){\makebox(0,0){0.2}}
\put(482,113){\special{em:moveto}}
\put(482,133){\special{em:lineto}}
\put(482,697){\special{em:moveto}}
\put(482,677){\special{em:lineto}}
\put(482,68){\makebox(0,0){0.4}}
\put(613,113){\special{em:moveto}}
\put(613,133){\special{em:lineto}}
\put(613,697){\special{em:moveto}}
\put(613,677){\special{em:lineto}}
\put(613,68){\makebox(0,0){0.6}}
\put(743,113){\special{em:moveto}}
\put(743,133){\special{em:lineto}}
\put(743,697){\special{em:moveto}}
\put(743,677){\special{em:lineto}}
\put(743,68){\makebox(0,0){0.8}}
\put(874,113){\special{em:moveto}}
\put(874,133){\special{em:lineto}}
\put(874,697){\special{em:moveto}}
\put(874,677){\special{em:lineto}}
\put(874,68){\makebox(0,0){1}}
\put(1005,113){\special{em:moveto}}
\put(1005,133){\special{em:lineto}}
\put(1005,697){\special{em:moveto}}
\put(1005,677){\special{em:lineto}}
\put(1005,68){\makebox(0,0){1.2}}
\put(1136,113){\special{em:moveto}}
\put(1136,133){\special{em:lineto}}
\put(1136,697){\special{em:moveto}}
\put(1136,677){\special{em:lineto}}
\put(1136,68){\makebox(0,0){1.4}}
\put(220,113){\special{em:moveto}}
\put(1136,113){\special{em:lineto}}
\put(1136,697){\special{em:lineto}}
\put(220,697){\special{em:lineto}}
\put(220,113){\special{em:lineto}}
\put(45,405){\makebox(0,0){$T/T_{BCS}^{MF}$}}
\put(678,23){\makebox(0,0){$\epsilon_{F}/ \omega_0$}}
\put(874,614){\makebox(0,0)[l]{$\alpha$=0}}
\sbox{\plotpoint}{\rule[-0.500pt]{1.000pt}{1.000pt}}%
\special{em:linewidth 1.0pt}%
\put(220,113){\usebox{\plotpoint}}
\multiput(220,113)(3.858,41.331){2}{\usebox{\plotpoint}}
\multiput(227,188)(3.936,41.324){2}{\usebox{\plotpoint}}
\put(236.80,278.17){\usebox{\plotpoint}}
\put(242.67,319.26){\usebox{\plotpoint}}
\put(249.74,360.15){\usebox{\plotpoint}}
\put(258.13,400.80){\usebox{\plotpoint}}
\multiput(259,405)(13.127,39.381){0}{\usebox{\plotpoint}}
\put(270.15,440.51){\usebox{\plotpoint}}
\multiput(272,447)(20.916,35.856){0}{\usebox{\plotpoint}}
\multiput(279,459)(17.396,37.690){0}{\usebox{\plotpoint}}
\put(287.63,478.00){\usebox{\plotpoint}}
\multiput(292,488)(33.779,24.128){0}{\usebox{\plotpoint}}
\multiput(299,493)(24.907,33.209){0}{\usebox{\plotpoint}}
\multiput(305,501)(36.042,20.595){0}{\usebox{\plotpoint}}
\put(315.27,507.18){\usebox{\plotpoint}}
\multiput(318,509)(36.042,20.595){0}{\usebox{\plotpoint}}
\multiput(325,513)(31.890,26.575){0}{\usebox{\plotpoint}}
\multiput(331,518)(41.094,5.871){0}{\usebox{\plotpoint}}
\multiput(338,519)(37.129,18.564){0}{\usebox{\plotpoint}}
\multiput(344,522)(41.094,5.871){0}{\usebox{\plotpoint}}
\put(352.35,523.67){\usebox{\plotpoint}}
\multiput(357,526)(41.094,5.871){0}{\usebox{\plotpoint}}
\multiput(364,527)(40.946,6.824){0}{\usebox{\plotpoint}}
\multiput(370,528)(39.914,11.404){0}{\usebox{\plotpoint}}
\multiput(377,530)(41.511,0.000){0}{\usebox{\plotpoint}}
\multiput(384,530)(41.511,0.000){0}{\usebox{\plotpoint}}
\put(392.88,530.00){\usebox{\plotpoint}}
\multiput(397,530)(41.511,0.000){0}{\usebox{\plotpoint}}
\multiput(403,530)(41.511,0.000){0}{\usebox{\plotpoint}}
\multiput(410,530)(41.511,0.000){0}{\usebox{\plotpoint}}
\multiput(416,530)(41.511,0.000){0}{\usebox{\plotpoint}}
\multiput(423,530)(41.511,0.000){0}{\usebox{\plotpoint}}
\put(434.39,530.00){\usebox{\plotpoint}}
\multiput(436,530)(41.511,0.000){0}{\usebox{\plotpoint}}
\multiput(442,530)(41.511,0.000){0}{\usebox{\plotpoint}}
\multiput(449,530)(41.511,0.000){0}{\usebox{\plotpoint}}
\multiput(456,530)(41.511,0.000){0}{\usebox{\plotpoint}}
\multiput(462,530)(41.511,0.000){0}{\usebox{\plotpoint}}
\multiput(469,530)(41.511,0.000){0}{\usebox{\plotpoint}}
\put(475.90,530.00){\usebox{\plotpoint}}
\multiput(482,530)(41.511,0.000){0}{\usebox{\plotpoint}}
\multiput(488,530)(41.511,0.000){0}{\usebox{\plotpoint}}
\multiput(495,530)(41.511,0.000){0}{\usebox{\plotpoint}}
\multiput(501,530)(41.511,0.000){0}{\usebox{\plotpoint}}
\multiput(508,530)(41.511,0.000){0}{\usebox{\plotpoint}}
\put(517.41,530.00){\usebox{\plotpoint}}
\multiput(521,530)(41.511,0.000){0}{\usebox{\plotpoint}}
\multiput(528,530)(41.511,0.000){0}{\usebox{\plotpoint}}
\multiput(534,530)(41.511,0.000){0}{\usebox{\plotpoint}}
\multiput(541,530)(41.511,0.000){0}{\usebox{\plotpoint}}
\multiput(547,530)(41.511,0.000){0}{\usebox{\plotpoint}}
\put(558.92,530.00){\usebox{\plotpoint}}
\multiput(560,530)(41.511,0.000){0}{\usebox{\plotpoint}}
\multiput(567,530)(41.511,0.000){0}{\usebox{\plotpoint}}
\multiput(573,530)(41.511,0.000){0}{\usebox{\plotpoint}}
\multiput(580,530)(41.511,0.000){0}{\usebox{\plotpoint}}
\multiput(586,530)(41.511,0.000){0}{\usebox{\plotpoint}}
\multiput(593,530)(41.511,0.000){0}{\usebox{\plotpoint}}
\put(600.43,530.00){\usebox{\plotpoint}}
\multiput(606,530)(41.511,0.000){0}{\usebox{\plotpoint}}
\multiput(613,530)(41.511,0.000){0}{\usebox{\plotpoint}}
\multiput(619,530)(41.511,0.000){0}{\usebox{\plotpoint}}
\multiput(626,530)(41.511,0.000){0}{\usebox{\plotpoint}}
\multiput(632,530)(41.511,0.000){0}{\usebox{\plotpoint}}
\put(641.94,530.00){\usebox{\plotpoint}}
\multiput(645,530)(41.511,0.000){0}{\usebox{\plotpoint}}
\multiput(652,530)(41.511,0.000){0}{\usebox{\plotpoint}}
\multiput(658,530)(41.511,0.000){0}{\usebox{\plotpoint}}
\multiput(665,530)(41.511,0.000){0}{\usebox{\plotpoint}}
\multiput(671,530)(41.511,0.000){0}{\usebox{\plotpoint}}
\put(683.45,530.00){\usebox{\plotpoint}}
\multiput(685,530)(41.511,0.000){0}{\usebox{\plotpoint}}
\multiput(691,530)(41.511,0.000){0}{\usebox{\plotpoint}}
\multiput(698,530)(41.511,0.000){0}{\usebox{\plotpoint}}
\multiput(704,530)(41.511,0.000){0}{\usebox{\plotpoint}}
\multiput(711,530)(41.511,0.000){0}{\usebox{\plotpoint}}
\multiput(717,530)(41.511,0.000){0}{\usebox{\plotpoint}}
\put(724.97,530.00){\usebox{\plotpoint}}
\multiput(730,530)(41.511,0.000){0}{\usebox{\plotpoint}}
\multiput(737,530)(41.511,0.000){0}{\usebox{\plotpoint}}
\multiput(743,530)(41.511,0.000){7}{\usebox{\plotpoint}}
\multiput(1038,530)(41.511,0.000){2}{\usebox{\plotpoint}}
\put(1136,530){\usebox{\plotpoint}}
\sbox{\plotpoint}{\rule[-0.200pt]{0.400pt}{0.400pt}}%
\special{em:linewidth 0.4pt}%
\put(220,113){\special{em:moveto}}
\put(233,114){\special{em:lineto}}
\put(246,116){\special{em:lineto}}
\put(259,117){\special{em:lineto}}
\put(272,119){\special{em:lineto}}
\put(285,120){\special{em:lineto}}
\put(299,121){\special{em:lineto}}
\put(312,123){\special{em:lineto}}
\put(325,124){\special{em:lineto}}
\put(338,125){\special{em:lineto}}
\put(351,127){\special{em:lineto}}
\put(364,128){\special{em:lineto}}
\put(377,129){\special{em:lineto}}
\put(390,130){\special{em:lineto}}
\put(403,131){\special{em:lineto}}
\put(416,132){\special{em:lineto}}
\put(429,134){\special{em:lineto}}
\put(482,138){\special{em:lineto}}
\put(547,143){\special{em:lineto}}
\put(613,149){\special{em:lineto}}
\put(678,153){\special{em:lineto}}
\put(743,158){\special{em:lineto}}
\put(809,162){\special{em:lineto}}
\put(874,166){\special{em:lineto}}
\put(940,169){\special{em:lineto}}
\put(1005,173){\special{em:lineto}}
\put(1077,177){\special{em:lineto}}
\put(1136,180){\special{em:lineto}}
\end{picture}\\
\setlength{\unitlength}{0.240900pt}
\ifx\plotpoint\undefined\newsavebox{\plotpoint}\fi
\sbox{\plotpoint}{\rule[-0.200pt]{0.400pt}{0.400pt}}%
\special{em:linewidth 0.4pt}%
\begin{picture}(1200,720)(0,0)
\font\gnuplot=cmr10 at 10pt
\gnuplot
\put(220,113){\special{em:moveto}}
\put(1136,113){\special{em:lineto}}
\put(220,113){\special{em:moveto}}
\put(220,697){\special{em:lineto}}
\put(220,113){\special{em:moveto}}
\put(240,113){\special{em:lineto}}
\put(1136,113){\special{em:moveto}}
\put(1116,113){\special{em:lineto}}
\put(198,113){\makebox(0,0)[r]{0}}
\put(220,530){\special{em:moveto}}
\put(240,530){\special{em:lineto}}
\put(1136,530){\special{em:moveto}}
\put(1116,530){\special{em:lineto}}
\put(198,530){\makebox(0,0)[r]{1}}
\put(220,113){\special{em:moveto}}
\put(220,133){\special{em:lineto}}
\put(220,697){\special{em:moveto}}
\put(220,677){\special{em:lineto}}
\put(220,68){\makebox(0,0){0}}
\put(351,113){\special{em:moveto}}
\put(351,133){\special{em:lineto}}
\put(351,697){\special{em:moveto}}
\put(351,677){\special{em:lineto}}
\put(351,68){\makebox(0,0){0.2}}
\put(482,113){\special{em:moveto}}
\put(482,133){\special{em:lineto}}
\put(482,697){\special{em:moveto}}
\put(482,677){\special{em:lineto}}
\put(482,68){\makebox(0,0){0.4}}
\put(613,113){\special{em:moveto}}
\put(613,133){\special{em:lineto}}
\put(613,697){\special{em:moveto}}
\put(613,677){\special{em:lineto}}
\put(613,68){\makebox(0,0){0.6}}
\put(743,113){\special{em:moveto}}
\put(743,133){\special{em:lineto}}
\put(743,697){\special{em:moveto}}
\put(743,677){\special{em:lineto}}
\put(743,68){\makebox(0,0){0.8}}
\put(874,113){\special{em:moveto}}
\put(874,133){\special{em:lineto}}
\put(874,697){\special{em:moveto}}
\put(874,677){\special{em:lineto}}
\put(874,68){\makebox(0,0){1}}
\put(1005,113){\special{em:moveto}}
\put(1005,133){\special{em:lineto}}
\put(1005,697){\special{em:moveto}}
\put(1005,677){\special{em:lineto}}
\put(1005,68){\makebox(0,0){1.2}}
\put(1136,113){\special{em:moveto}}
\put(1136,133){\special{em:lineto}}
\put(1136,697){\special{em:moveto}}
\put(1136,677){\special{em:lineto}}
\put(1136,68){\makebox(0,0){1.4}}
\put(220,113){\special{em:moveto}}
\put(1136,113){\special{em:lineto}}
\put(1136,697){\special{em:lineto}}
\put(220,697){\special{em:lineto}}
\put(220,113){\special{em:lineto}}
\put(45,405){\makebox(0,0){$T/T_{BCS}^{MF}$}}
\put(678,23){\makebox(0,0){$\epsilon_{F}/ \omega_0$}}
\put(874,614){\makebox(0,0)[l]{$\alpha$=0.1}}
\sbox{\plotpoint}{\rule[-0.500pt]{1.000pt}{1.000pt}}%
\special{em:linewidth 1.0pt}%
\put(220,113){\usebox{\plotpoint}}
\multiput(220,113)(5.687,41.120){5}{\usebox{\plotpoint}}
\put(248.99,318.51){\usebox{\plotpoint}}
\put(256.07,359.41){\usebox{\plotpoint}}
\put(264.79,399.98){\usebox{\plotpoint}}
\multiput(266,405)(11.404,39.914){0}{\usebox{\plotpoint}}
\put(276.59,439.76){\usebox{\plotpoint}}
\multiput(279,447)(18.564,37.129){0}{\usebox{\plotpoint}}
\multiput(285,459)(19.680,36.549){0}{\usebox{\plotpoint}}
\put(294.28,477.22){\usebox{\plotpoint}}
\multiput(299,488)(31.890,26.575){0}{\usebox{\plotpoint}}
\multiput(305,493)(27.335,31.240){0}{\usebox{\plotpoint}}
\multiput(312,501)(34.539,23.026){0}{\usebox{\plotpoint}}
\put(321.56,507.03){\usebox{\plotpoint}}
\multiput(325,509)(34.539,23.026){0}{\usebox{\plotpoint}}
\multiput(331,513)(33.779,24.128){0}{\usebox{\plotpoint}}
\multiput(338,518)(40.946,6.824){0}{\usebox{\plotpoint}}
\multiput(344,519)(38.155,16.352){0}{\usebox{\plotpoint}}
\multiput(351,522)(40.946,6.824){0}{\usebox{\plotpoint}}
\put(358.79,523.77){\usebox{\plotpoint}}
\multiput(364,526)(40.946,6.824){0}{\usebox{\plotpoint}}
\multiput(370,527)(41.094,5.871){0}{\usebox{\plotpoint}}
\multiput(377,528)(39.914,11.404){0}{\usebox{\plotpoint}}
\multiput(384,530)(41.511,0.000){0}{\usebox{\plotpoint}}
\multiput(390,530)(41.511,0.000){0}{\usebox{\plotpoint}}
\put(399.41,530.00){\usebox{\plotpoint}}
\multiput(403,530)(41.511,0.000){0}{\usebox{\plotpoint}}
\multiput(410,530)(41.511,0.000){0}{\usebox{\plotpoint}}
\multiput(416,530)(41.511,0.000){0}{\usebox{\plotpoint}}
\multiput(423,530)(41.511,0.000){0}{\usebox{\plotpoint}}
\multiput(429,530)(41.511,0.000){0}{\usebox{\plotpoint}}
\put(440.92,530.00){\usebox{\plotpoint}}
\multiput(442,530)(41.511,0.000){0}{\usebox{\plotpoint}}
\multiput(449,530)(41.511,0.000){0}{\usebox{\plotpoint}}
\multiput(456,530)(41.511,0.000){0}{\usebox{\plotpoint}}
\multiput(462,530)(41.511,0.000){0}{\usebox{\plotpoint}}
\multiput(469,530)(41.511,0.000){0}{\usebox{\plotpoint}}
\multiput(475,530)(41.511,0.000){0}{\usebox{\plotpoint}}
\put(482.43,530.00){\usebox{\plotpoint}}
\multiput(488,530)(41.511,0.000){0}{\usebox{\plotpoint}}
\multiput(495,530)(41.511,0.000){0}{\usebox{\plotpoint}}
\multiput(501,530)(41.511,0.000){0}{\usebox{\plotpoint}}
\multiput(508,530)(41.511,0.000){0}{\usebox{\plotpoint}}
\multiput(514,530)(41.511,0.000){0}{\usebox{\plotpoint}}
\put(523.94,530.00){\usebox{\plotpoint}}
\multiput(528,530)(41.511,0.000){0}{\usebox{\plotpoint}}
\multiput(534,530)(41.511,0.000){0}{\usebox{\plotpoint}}
\multiput(541,530)(41.511,0.000){0}{\usebox{\plotpoint}}
\multiput(547,530)(41.511,0.000){0}{\usebox{\plotpoint}}
\multiput(554,530)(41.511,0.000){0}{\usebox{\plotpoint}}
\put(565.46,530.00){\usebox{\plotpoint}}
\multiput(567,530)(41.511,0.000){0}{\usebox{\plotpoint}}
\multiput(573,530)(41.511,0.000){0}{\usebox{\plotpoint}}
\multiput(580,530)(41.511,0.000){0}{\usebox{\plotpoint}}
\multiput(586,530)(41.511,0.000){0}{\usebox{\plotpoint}}
\multiput(593,530)(41.511,0.000){0}{\usebox{\plotpoint}}
\multiput(599,530)(41.511,0.000){0}{\usebox{\plotpoint}}
\put(606.97,530.00){\usebox{\plotpoint}}
\multiput(613,530)(41.511,0.000){0}{\usebox{\plotpoint}}
\multiput(619,530)(41.511,0.000){0}{\usebox{\plotpoint}}
\multiput(626,530)(41.511,0.000){0}{\usebox{\plotpoint}}
\multiput(632,530)(41.511,0.000){0}{\usebox{\plotpoint}}
\multiput(639,530)(41.511,0.000){0}{\usebox{\plotpoint}}
\put(648.48,530.00){\usebox{\plotpoint}}
\multiput(652,530)(41.511,0.000){0}{\usebox{\plotpoint}}
\multiput(658,530)(41.511,0.000){0}{\usebox{\plotpoint}}
\multiput(665,530)(41.511,0.000){0}{\usebox{\plotpoint}}
\multiput(671,530)(41.511,0.000){0}{\usebox{\plotpoint}}
\multiput(678,530)(41.511,0.000){0}{\usebox{\plotpoint}}
\put(689.99,530.00){\usebox{\plotpoint}}
\multiput(691,530)(41.511,0.000){0}{\usebox{\plotpoint}}
\multiput(698,530)(41.511,0.000){0}{\usebox{\plotpoint}}
\multiput(704,530)(41.511,0.000){0}{\usebox{\plotpoint}}
\multiput(711,530)(41.511,0.000){0}{\usebox{\plotpoint}}
\multiput(717,530)(41.511,0.000){0}{\usebox{\plotpoint}}
\multiput(724,530)(41.511,0.000){0}{\usebox{\plotpoint}}
\put(731.50,530.00){\usebox{\plotpoint}}
\multiput(737,530)(41.511,0.000){0}{\usebox{\plotpoint}}
\multiput(743,530)(41.511,0.000){7}{\usebox{\plotpoint}}
\multiput(1038,530)(41.511,0.000){2}{\usebox{\plotpoint}}
\put(1136,530){\usebox{\plotpoint}}
\sbox{\plotpoint}{\rule[-0.200pt]{0.400pt}{0.400pt}}%
\special{em:linewidth 0.4pt}%
\put(220,113){\special{em:moveto}}
\put(233,114){\special{em:lineto}}
\put(246,115){\special{em:lineto}}
\put(259,117){\special{em:lineto}}
\put(272,118){\special{em:lineto}}
\put(285,119){\special{em:lineto}}
\put(299,121){\special{em:lineto}}
\put(312,122){\special{em:lineto}}
\put(325,123){\special{em:lineto}}
\put(338,125){\special{em:lineto}}
\put(351,126){\special{em:lineto}}
\put(364,127){\special{em:lineto}}
\put(377,128){\special{em:lineto}}
\put(390,130){\special{em:lineto}}
\put(403,131){\special{em:lineto}}
\put(416,132){\special{em:lineto}}
\put(429,133){\special{em:lineto}}
\put(482,138){\special{em:lineto}}
\put(547,143){\special{em:lineto}}
\put(613,148){\special{em:lineto}}
\put(678,153){\special{em:lineto}}
\put(743,157){\special{em:lineto}}
\put(809,161){\special{em:lineto}}
\put(874,165){\special{em:lineto}}
\put(940,169){\special{em:lineto}}
\put(1005,173){\special{em:lineto}}
\put(1077,176){\special{em:lineto}}
\put(1136,180){\special{em:lineto}}
\end{picture}\\
\setlength{\unitlength}{0.240900pt}
\ifx\plotpoint\undefined\newsavebox{\plotpoint}\fi
\sbox{\plotpoint}{\rule[-0.200pt]{0.400pt}{0.400pt}}%
\special{em:linewidth 0.4pt}%
\begin{picture}(1200,720)(0,0)
\font\gnuplot=cmr10 at 10pt
\gnuplot
\put(220,113){\special{em:moveto}}
\put(1136,113){\special{em:lineto}}
\put(220,113){\special{em:moveto}}
\put(220,697){\special{em:lineto}}
\put(220,113){\special{em:moveto}}
\put(240,113){\special{em:lineto}}
\put(1136,113){\special{em:moveto}}
\put(1116,113){\special{em:lineto}}
\put(198,113){\makebox(0,0)[r]{0}}
\put(220,530){\special{em:moveto}}
\put(240,530){\special{em:lineto}}
\put(1136,530){\special{em:moveto}}
\put(1116,530){\special{em:lineto}}
\put(198,530){\makebox(0,0)[r]{1}}
\put(220,113){\special{em:moveto}}
\put(220,133){\special{em:lineto}}
\put(220,697){\special{em:moveto}}
\put(220,677){\special{em:lineto}}
\put(220,68){\makebox(0,0){0}}
\put(351,113){\special{em:moveto}}
\put(351,133){\special{em:lineto}}
\put(351,697){\special{em:moveto}}
\put(351,677){\special{em:lineto}}
\put(351,68){\makebox(0,0){0.2}}
\put(482,113){\special{em:moveto}}
\put(482,133){\special{em:lineto}}
\put(482,697){\special{em:moveto}}
\put(482,677){\special{em:lineto}}
\put(482,68){\makebox(0,0){0.4}}
\put(613,113){\special{em:moveto}}
\put(613,133){\special{em:lineto}}
\put(613,697){\special{em:moveto}}
\put(613,677){\special{em:lineto}}
\put(613,68){\makebox(0,0){0.6}}
\put(743,113){\special{em:moveto}}
\put(743,133){\special{em:lineto}}
\put(743,697){\special{em:moveto}}
\put(743,677){\special{em:lineto}}
\put(743,68){\makebox(0,0){0.8}}
\put(874,113){\special{em:moveto}}
\put(874,133){\special{em:lineto}}
\put(874,697){\special{em:moveto}}
\put(874,677){\special{em:lineto}}
\put(874,68){\makebox(0,0){1}}
\put(1005,113){\special{em:moveto}}
\put(1005,133){\special{em:lineto}}
\put(1005,697){\special{em:moveto}}
\put(1005,677){\special{em:lineto}}
\put(1005,68){\makebox(0,0){1.2}}
\put(1136,113){\special{em:moveto}}
\put(1136,133){\special{em:lineto}}
\put(1136,697){\special{em:moveto}}
\put(1136,677){\special{em:lineto}}
\put(1136,68){\makebox(0,0){1.4}}
\put(220,113){\special{em:moveto}}
\put(1136,113){\special{em:lineto}}
\put(1136,697){\special{em:lineto}}
\put(220,697){\special{em:lineto}}
\put(220,113){\special{em:lineto}}
\put(45,405){\makebox(0,0){$T/T_{BCS}^{MF}$}}
\put(678,23){\makebox(0,0){$\epsilon_{F}/ \omega_0$}}
\put(874,614){\makebox(0,0)[l]{$\alpha$=1.1}}
\sbox{\plotpoint}{\rule[-0.500pt]{1.000pt}{1.000pt}}%
\special{em:linewidth 1.0pt}%
\put(220,113){\usebox{\plotpoint}}
\multiput(220,113)(5.687,41.120){5}{\usebox{\plotpoint}}
\put(248.99,318.51){\usebox{\plotpoint}}
\put(256.07,359.41){\usebox{\plotpoint}}
\put(264.79,399.98){\usebox{\plotpoint}}
\multiput(266,405)(11.404,39.914){0}{\usebox{\plotpoint}}
\put(276.59,439.76){\usebox{\plotpoint}}
\multiput(279,447)(18.564,37.129){0}{\usebox{\plotpoint}}
\multiput(285,459)(19.680,36.549){0}{\usebox{\plotpoint}}
\put(294.28,477.22){\usebox{\plotpoint}}
\multiput(299,488)(31.890,26.575){0}{\usebox{\plotpoint}}
\multiput(305,493)(27.335,31.240){0}{\usebox{\plotpoint}}
\multiput(312,501)(34.539,23.026){0}{\usebox{\plotpoint}}
\put(321.56,507.03){\usebox{\plotpoint}}
\multiput(325,509)(34.539,23.026){0}{\usebox{\plotpoint}}
\multiput(331,513)(33.779,24.128){0}{\usebox{\plotpoint}}
\multiput(338,518)(40.946,6.824){0}{\usebox{\plotpoint}}
\multiput(344,519)(38.155,16.352){0}{\usebox{\plotpoint}}
\multiput(351,522)(40.946,6.824){0}{\usebox{\plotpoint}}
\put(358.79,523.77){\usebox{\plotpoint}}
\multiput(364,526)(40.946,6.824){0}{\usebox{\plotpoint}}
\multiput(370,527)(41.094,5.871){0}{\usebox{\plotpoint}}
\multiput(377,528)(39.914,11.404){0}{\usebox{\plotpoint}}
\multiput(384,530)(41.511,0.000){0}{\usebox{\plotpoint}}
\multiput(390,530)(41.511,0.000){0}{\usebox{\plotpoint}}
\put(399.41,530.00){\usebox{\plotpoint}}
\multiput(403,530)(41.511,0.000){0}{\usebox{\plotpoint}}
\multiput(410,530)(41.511,0.000){0}{\usebox{\plotpoint}}
\multiput(416,530)(41.511,0.000){0}{\usebox{\plotpoint}}
\multiput(423,530)(41.511,0.000){0}{\usebox{\plotpoint}}
\multiput(429,530)(41.511,0.000){0}{\usebox{\plotpoint}}
\put(440.92,530.00){\usebox{\plotpoint}}
\multiput(442,530)(41.511,0.000){0}{\usebox{\plotpoint}}
\multiput(449,530)(41.511,0.000){0}{\usebox{\plotpoint}}
\multiput(456,530)(41.511,0.000){0}{\usebox{\plotpoint}}
\multiput(462,530)(41.511,0.000){0}{\usebox{\plotpoint}}
\multiput(469,530)(41.511,0.000){0}{\usebox{\plotpoint}}
\multiput(475,530)(41.511,0.000){0}{\usebox{\plotpoint}}
\put(482.43,530.00){\usebox{\plotpoint}}
\multiput(488,530)(41.511,0.000){0}{\usebox{\plotpoint}}
\multiput(495,530)(41.511,0.000){0}{\usebox{\plotpoint}}
\multiput(501,530)(41.511,0.000){0}{\usebox{\plotpoint}}
\multiput(508,530)(41.511,0.000){0}{\usebox{\plotpoint}}
\multiput(514,530)(41.511,0.000){0}{\usebox{\plotpoint}}
\put(523.94,530.00){\usebox{\plotpoint}}
\multiput(528,530)(41.511,0.000){0}{\usebox{\plotpoint}}
\multiput(534,530)(41.511,0.000){0}{\usebox{\plotpoint}}
\multiput(541,530)(41.511,0.000){0}{\usebox{\plotpoint}}
\multiput(547,530)(41.511,0.000){0}{\usebox{\plotpoint}}
\multiput(554,530)(41.511,0.000){0}{\usebox{\plotpoint}}
\put(565.46,530.00){\usebox{\plotpoint}}
\multiput(567,530)(41.511,0.000){0}{\usebox{\plotpoint}}
\multiput(573,530)(41.511,0.000){0}{\usebox{\plotpoint}}
\multiput(580,530)(41.511,0.000){0}{\usebox{\plotpoint}}
\multiput(586,530)(41.511,0.000){0}{\usebox{\plotpoint}}
\multiput(593,530)(41.511,0.000){0}{\usebox{\plotpoint}}
\multiput(599,530)(41.511,0.000){0}{\usebox{\plotpoint}}
\put(606.97,530.00){\usebox{\plotpoint}}
\multiput(613,530)(41.511,0.000){0}{\usebox{\plotpoint}}
\multiput(619,530)(41.511,0.000){0}{\usebox{\plotpoint}}
\multiput(626,530)(41.511,0.000){0}{\usebox{\plotpoint}}
\multiput(632,530)(41.511,0.000){0}{\usebox{\plotpoint}}
\multiput(639,530)(41.511,0.000){0}{\usebox{\plotpoint}}
\put(648.48,530.00){\usebox{\plotpoint}}
\multiput(652,530)(41.511,0.000){0}{\usebox{\plotpoint}}
\multiput(658,530)(41.511,0.000){0}{\usebox{\plotpoint}}
\multiput(665,530)(41.511,0.000){0}{\usebox{\plotpoint}}
\multiput(671,530)(41.511,0.000){0}{\usebox{\plotpoint}}
\multiput(678,530)(41.511,0.000){0}{\usebox{\plotpoint}}
\put(689.99,530.00){\usebox{\plotpoint}}
\multiput(691,530)(41.511,0.000){0}{\usebox{\plotpoint}}
\multiput(698,530)(41.511,0.000){0}{\usebox{\plotpoint}}
\multiput(704,530)(41.511,0.000){0}{\usebox{\plotpoint}}
\multiput(711,530)(41.511,0.000){0}{\usebox{\plotpoint}}
\multiput(717,530)(41.511,0.000){0}{\usebox{\plotpoint}}
\multiput(724,530)(41.511,0.000){0}{\usebox{\plotpoint}}
\put(731.50,530.00){\usebox{\plotpoint}}
\multiput(737,530)(41.511,0.000){0}{\usebox{\plotpoint}}
\multiput(743,530)(41.511,0.000){7}{\usebox{\plotpoint}}
\multiput(1038,530)(34.539,-23.026){0}{\usebox{\plotpoint}}
\multiput(1044,526)(36.042,-20.595){0}{\usebox{\plotpoint}}
\put(1055.32,512.64){\usebox{\plotpoint}}
\put(1063.97,472.15){\usebox{\plotpoint}}
\multiput(1064,472)(3.019,-41.401){2}{\usebox{\plotpoint}}
\multiput(1071,376)(0.947,-41.500){6}{\usebox{\plotpoint}}
\put(1077,113){\usebox{\plotpoint}}
\sbox{\plotpoint}{\rule[-0.200pt]{0.400pt}{0.400pt}}%
\special{em:linewidth 0.4pt}%
\put(220,113){\special{em:moveto}}
\put(233,114){\special{em:lineto}}
\put(246,116){\special{em:lineto}}
\put(259,117){\special{em:lineto}}
\put(272,119){\special{em:lineto}}
\put(285,120){\special{em:lineto}}
\put(299,121){\special{em:lineto}}
\put(312,123){\special{em:lineto}}
\put(325,124){\special{em:lineto}}
\put(338,125){\special{em:lineto}}
\put(351,127){\special{em:lineto}}
\put(364,128){\special{em:lineto}}
\put(377,129){\special{em:lineto}}
\put(390,130){\special{em:lineto}}
\put(403,131){\special{em:lineto}}
\put(416,132){\special{em:lineto}}
\put(429,134){\special{em:lineto}}
\put(482,138){\special{em:lineto}}
\put(547,143){\special{em:lineto}}
\put(613,149){\special{em:lineto}}
\put(678,153){\special{em:lineto}}
\put(743,158){\special{em:lineto}}
\put(809,162){\special{em:lineto}}
\put(874,166){\special{em:lineto}}
\put(940,169){\special{em:lineto}}
\put(1005,172){\special{em:lineto}}
\put(1057,175){\special{em:lineto}}
\put(1064,168){\special{em:lineto}}
\put(1071,151){\special{em:lineto}}
\put(1077,113){\special{em:lineto}}
\end{picture}\\
\end{centering}
\caption{The characteristic patterns
of the $T-n_{f}$ phase diagram of 2D metal with coupling
constant $\lambda =1$. Doted and solid lines
define the temperatures $T_{\rho}$ and $T_{BKT}$, respectively.
}
\end{figure}
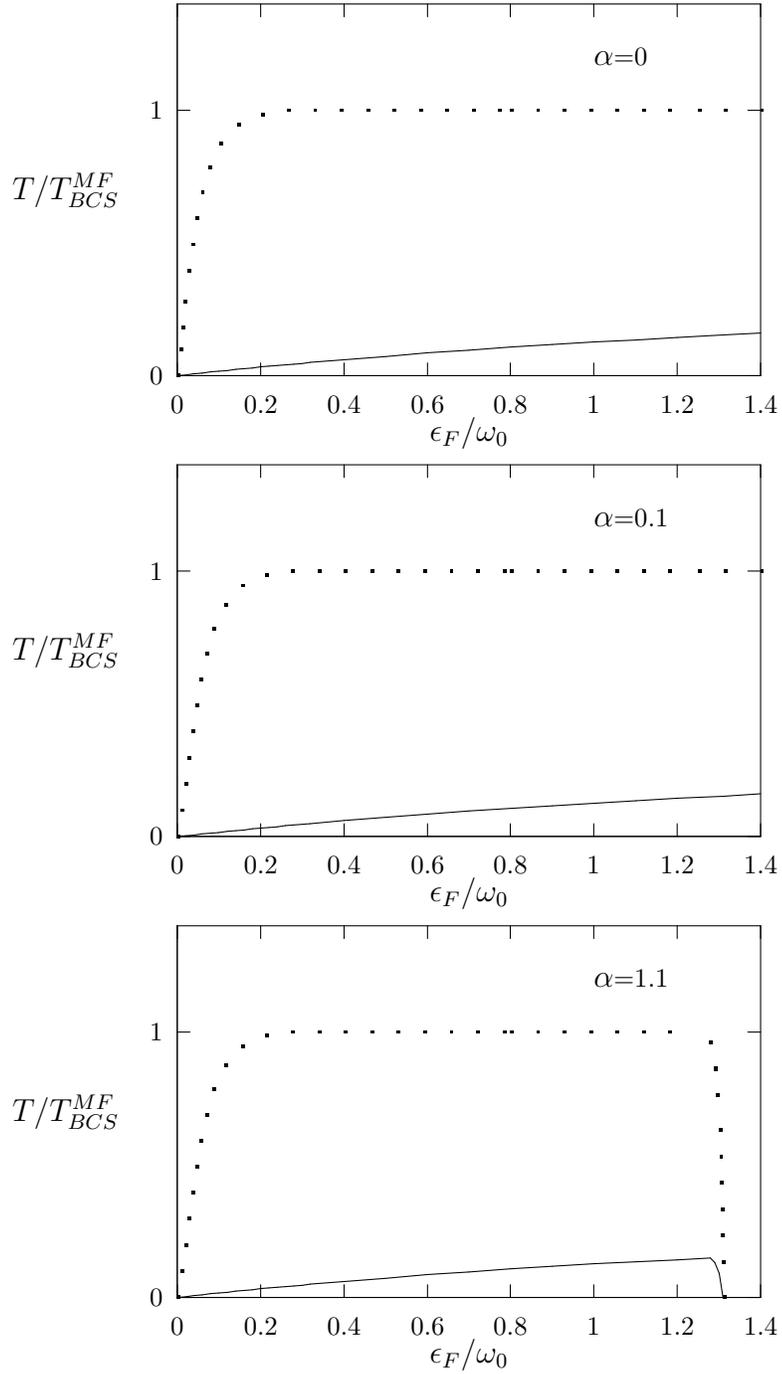

It is seen that due to the long-wave phonon
suppression (in
fact, to be supposed strong) the functions $T_{\rho}(n_{f})$ and
$T_{BKT}(n_{f})$ quickly on $n_{f}$ variation acquire a bell-like
shape.  This "bell" proves to be non-symmetrical, and its heigt,
width and to some extent form depend on $\alpha (>1)$.  In some sense
such a result is surprising because, as it is generally accepted, not
long-, but short-wave intermediate bosons play the main role in
attraction which appears in BCS-Eliashberg model due to
electron-phonon (or any other boson) interaction.  The sensitivity of
superconducting properties of a 2D metal to long-wave part of the
intermediate boson spectrum is rather unusual and
allows to hope the more accurate consideration also results in a
similar effect.

\section{Conclusion}

The existence of two-different temperatures in underdoped HTSCs is
now a well established fact. They, as it shown in many papers
(see the rewiev \cite{Lok4}), are one of the consequences of
two-dimensionality of their electronic and magnetic properties when
the "ordering" of the order parameter modulus and phase takes place
at different temperatures. In pure 2D system the temperature
$T_{BKT}$, as was mentioned, has to be considered as critical one, and in the
region $T_{BKT}<T<T_{\rho}$ so called pseudogap,
also normal, phase is formed in underdoped HTSCs. It is destroyed
when $T>T_{\rho}$ or $n_{f}$ becomes so large (overdoped regime) that
chemical potential of fermions and Fermi energy are indistinctive
($\mu =\epsilon_{F}$). And the aim of this paper was to demonstrate
that in the superconducting system with indirect attraction the role
of long-wave bosons (for instance, phonons, spin fluctuations and so
on) can be crucial in the region where standard Fermi-liquid theory
becomes already applicable.

In spite of some qualitative similarity between experimental
and obtained pictures the model considered is so simple (and even
rough because the propagator (2) was, in fact, postulated while
it must contain the boson damping only) that
any quantitative its use is almost senseless.  Therefore it must be
generalized on the taking into account such HTSC features as:
quasi-two-dimensionality (which results in real $T_{c}$),
intermediate boson dispersion $\omega ({\bf k})$ (as it takes place
for spin fluctuations) and damping $\gamma ({\bf k})$; pairing
anisotropy, etc. These problems will be considered separately.

$$
{\bf Acknolegements}
$$

We are pleased to devote this paper to well-known ukrainian theorist,
Corresponding member of the National Academy  of Sciences of Ukraine,
Prof.~I.V.~Stasyuk whose numerical remarks and critics are always
interesting, useful and helpful for us.


\begin{thebibliography}{99}

\bibitem{Lok1}
Loktev~V.M. Mechanisms of high-$T_{c}$ superconductivity
of copper oxides.// Low Temp. Phys., 1996, vol.22 , No 1, p.
1-32.
\bibitem{Dess}
Shen~Z.-X., Dessau~D.S. Electronic structure and photoemission
studies of late transition - metal oxides - Mott insulators and
high-temperature super-conductors.//
Phys.~Rep., 1995, vol.253, No 1-3, p.1-162.
\bibitem{Keim}
Keimer~B., Belk~N., Birgeneau~R.J.,
Cassano~A., Chen~C.Y., Greven~M., Kastner~M.A., Shirane~G.
Magnetic excitations in pure, lightly doped and weakly metallic
$La_{2}CuO_{4}$.//
Phys.  Rev. B, 1992, vol.46, No 21, p.14034-14053.
\bibitem{Tak}
Takagi~H.,
Cava~R.J., Marezio~M., Batlogg~B., Krajewski~J.J., Peck~W.F.,
Jr., Bordet~P., Cox~D.E. Disappearance of superconductivity in
overdoped
$La_{2-x}Sr_{x}CuO_{4}$ at structural phase boundary.//
Phys.  Rev. Lett., 1992, vol.68, No 25, p.3777-3780.
\bibitem{Nag}
Nagano~T., Tamioka~Y., Nakayama~Y., Kishio~K.,
Kitazawa~K. Bulk superconductivity in both tetragonal and
orthorhombic solid solutions of
$(La_{1-x}Sr_{x})_{2}CuO_{4-\delta}$ at structural phase boundary.//
Phys. Rev. B  1993, vol.48, No 13, p.9689-9696.
\bibitem{Uem}
Uemura~Y.J., Le~L.P., Luke~J.M., Sternlieb~B,J.,
Wu~W.D., Brewer~J.H., Riesman~T.M., Seaman~C.L., Maple~M.B.,
Ishikawa~M., Hinks~D.G., Jorgensen~J.D., Saito~G., Yamchi~H.
Basic similarities among cuprate, bismthate, organic, Chevrel-
Phase and heavy-fermion superconductors shown by penetration-
depth measurements.//  Phys.
Rev.  Lett.  1991, vol. 66, No20, p.2665-2668.
\bibitem{Yam}
Yamada~K., Wakimoto~S., Shirane~G., Lee~C.H.,
Kastner~M.A., Hosoya~S., Greven~M., Endoh~Y.,
Birgeneau~R.J. Direct observation of a magnetic gap in superconducting
$La_{1.85}Sr_{0.15}CuO_{4}$ ($T_{c}=37.3$K).//
Rev.  Lett.,  1995, vol. 75, No 8, p.1626-1629.
\bibitem{Emer}
Emery~V., Kivelson~S.A.
Importance of phase fluctuations in superconductors with small
 superfluid density.// Nature 1995, vol. 374, p.434-437.
\bibitem{Gus}
Gusynin~V.P., Loktev~V.M., Sharapov~S.G.
Phase diagram of 2D metal with variable number of carriers.//
JETP Lett., 1997, vol. 65, No 2, p.182-188; On peculiarities
 of superconducting state formation in 2D metallic systems.//
Low Temp. Phys., 1997, vol.23 , No 8, p.
\bibitem{Abr}
Abrikosov~A.A. Size of BCS ratio in strongly underdoped
high-$T_{c}$ cuprates.// Phys.  Rev. B 1997, vol. 55, No 10,
p.R 6149-R 6151.
\bibitem{Lok4}
Loktev~V.M.,
Sharapov~S.G. Superconducting condensate formation in metallic
systems with arbitrary carrier density.// Cond. Mat. Phys., 1997,
No 11, p.131-178.
\bibitem{Plak}
Plakida~N.M. High-Temperature Superconductors, Berlin, Springer
Verlag, 1995.
\bibitem{St1}
Plakida~N.M., Yushankhai~V.Yu,
Stasyuk~I.V. On the role of kinematic and exchange
interactions in superconducting pairing of electrons in the
Hubbard model.// Physica C, 1989, vol. 160, No 1, p.80-88.
\bibitem{St2}
Stasyuk~I.V., Shvaika~A.M., Schachinger~E. On the electron spectrum
of the Hubbard model including the interaction with loca anharmonic
vibrations.//
Physica C, 1993, vol. 213, p.57-70.
\bibitem{St3}
Stasyuk~I.V., Shvaika~A.M. Dielectric properties and electron
spectrum of the M\" uller model in HTSC theory.//
 Acta Phys.Polonica A, 1993, vol. 84, No 2, p.293-313.
 \bibitem{St4}
Stasyuk~I.V., Shvaika~A.M., Schachinger~E. Electron spectrum
of the Hubbard model with coupling to pseudospin degrees of
freedom.//
Physica B, 1994, vol. 194-196, p.1965-1966.
\bibitem{Fried}
Friedel~J. The high-$T_{c}$ superconductors: a conservative view.//
J.Phys.:  Cond.  Mat., 1989, vol. 1, No 42, p. 7757-7794.
\bibitem{Mar} Markiewicz~R.S. Van Hove singularities and
high-$T_{c}$ superconductivity: A review.// Int. J.  Mod.
Phys.  B, 1991, vol.5, No 12, p. 2037-2071.
\bibitem{Gin}
Ginzburg~V.L. Superconductivity and
superfluidity (what is done and what is not done).// Usp. Fiz. Nauk,
1997, vol. 197, No 4,
p.429-454 (in Russian).
\bibitem{Kam}
Kampf~A.P. Magnetic correlations in high-temperature
superconductivity.// Phys.  Rep., 1994, vol. 249, No 4,5,
p.219-351.  \bibitem{Pin} Pines~D. Understanding high temperature
superconductivity:  progress and prospects.// CNLS Newsletter, Los
Alamos Nat. Lab., 1997, No 138.
\bibitem{Pash}
Pashitskii~E.A. Low
frequency charge excitations in cuprate metal-oxide compounds// Fiz.
Nizkikh Temp., 1995, vol. 21, No 10, p.995-1019; No 11, p.1091-1137
(in Russian).
\bibitem{Ivan}
Ivanov~M.A., Loktev~V.M.,
Pogorelov~Yu.G.  Localization of spin excitations and disruption of
long-range order in weakly doped $La_{2}CuO_{4}$.//JETP, 1992,
vol.74, No 2, p.317-325.
\bibitem{Lok2}
Loktev~V.M. About
correlation length in magnetic subsystem of high-$T_{c}$
superconductors.// Supercond.:  Physics, Chemistry, Technics, 1991,
vol. 4, No 12, p.2993-2296.
\bibitem{Thur}
Thurston~T.R.,
Birgeneau~R.J., Gabbe~D.R., Jenssen~H.P., Kastner~M.A., Picone~P.J.,
Preyer~N.W., Axe~J.D., B\" oni~P., Shirane~G., Sato~M., Fukuda~K.,
Shamoto~S. Neutron scattering study of soft optical phonons in
$La_{2-x}Sr_{x}CuO_{4-y}$.// Phys.  Rev.  B, 1989, vol. 39, No 7,
p.4327-4333.  \bibitem{Lok3} Loktev~V.M., Pogorelov~Yu.G.
Metallization and superconductivity in doped metal-oxide layered
compounds.// Physica C, 1996, vol. 272, p.151-160.  \bibitem{Cor1}
Cornwell~J.M., Jackiw~R., Tomboulis~E.
Effective action for composite operators
// Phys.~Rev.~D, 1974, vol. 10, No 8, P.2428-2445.
\bibitem{Lv}
Loktev~V.M., Sharapov~S.G., Turkowski~V.M.
Crossover from supefluidity to superconductivity
in 2D systems with indirect inter-carrier interaction.//
J. of Phys. Studies, 1997, vol. 1, No 3, p.431--440.
\bibitem{Iz}
Izyumov~Yu.A., Skryabin~Yu.N. Statistical mechanics of magnetically
ordered systems. Moscow, Nauka, 1987 (in Russian).
\end{thebibliography}
\end{document}